\documentclass[aps,reprint,twocolumn,superscriptaddress,citeautoscript,showpacs,amsmath,amssymb,prb]{revtex4-2}
\usepackage{amsmath}
\usepackage{graphicx}
\usepackage{dcolumn}
\usepackage{bm}
\usepackage{times}
\usepackage{color}
\usepackage{braket}
\usepackage{bm}
\begin{document}
\title{Short-range magnetic correlations in quasicrystalline \textit{i}-Tb-Cd}

\author{P. Das}
\altaffiliation{Present Address: Dynamic Compression Sector, Institute for Shock Physics, Washington State University, Argonne, Illinois 60439, USA}
\affiliation{Department of Physics and Astronomy, Iowa State University and Ames National Laboratory, Ames, Iowa 50011, USA}

\author{A. Kreyssig}
\affiliation{Department of Physics and Astronomy, Iowa State University and Ames National Laboratory, Ames, Iowa 50011, USA}

\author{G. S. Tucker}
\affiliation{Department of Physics and Astronomy, Iowa State University and Ames National Laboratory, Ames, Iowa 50011, USA}

\author{A. Podlesnyak}
\affiliation{Neutron Scattering Division, Oak Ridge National Laboratory, Oak Ridge, Tennessee 37831, USA}

\author{Feng Ye}
\affiliation{Neutron Scattering Division, Oak Ridge National Laboratory, Oak Ridge, Tennessee 37831, USA}

\author{Masaaki Matsuda}
\affiliation{Neutron Scattering Division, Oak Ridge National Laboratory, Oak Ridge, Tennessee 37831, USA}

\author{T. Kong}
\altaffiliation{Present Address: Department of Physics, University of Arizona, Tucson, AZ 85721, USA}
\affiliation{Department of Physics and Astronomy, Iowa State University and Ames National Laboratory, Ames, Iowa 50011, USA}

\author{S. L. Bud'ko}
\affiliation{Department of Physics and Astronomy, Iowa State University and Ames National Laboratory, Ames, Iowa 50011, USA}

\author{P. C. Canfield}
\affiliation{Department of Physics and Astronomy, Iowa State University and Ames National Laboratory, Ames, Iowa 50011, USA}

\author{R. Flint}
\affiliation{Department of Physics and Astronomy, Iowa State University and Ames National Laboratory, Ames, Iowa 50011, USA}

\author{P. P. Orth}
\altaffiliation{Present Address: Department of Physics, Saarland University, 66123 Saarbr\"ucken, Germany}
\affiliation{Department of Physics and Astronomy, Iowa State University and Ames National Laboratory, Ames, Iowa 50011, USA}
\affiliation{Department of Physics, Harvard University, Cambridge, Massachusetts 02138, USA}

\author{T. Yamada}
\affiliation{Department of Applied Physics, Tokyo University of Science, 6-3-1 Nijuku, Katsushika, Tokyo 125-8585, Japan}

\author{R. J. McQueeney}
\affiliation{Department of Physics and Astronomy, Iowa State University and Ames National Laboratory, Ames, Iowa 50011, USA}

\author{A. I. Goldman}
\email{email: goldman@iastate.edu} 
\affiliation{Department of Physics and Astronomy, Iowa State University and Ames National Laboratory, Ames, Iowa 50011, USA}

\date{\today}

\begin{abstract}
We report on elastic and inelastic neutron scattering from single-grain isotopically-enriched samples to elucidate the local magnetic correlations between Tb$^{3+}$ moments in quasicrystalline \textit{i}-Tb-Cd.  The inelastic neutron scattering measurements of the CEF excitations demonstrated that the Tb$^{3+}$ moments are directed primarily along the local five-fold axes of the Tsai-type cluster as was found for the TbCd$_6$ approximant phase. Based on the inelastic measurements, we consider of a simple Ising-type model for the moment configurations on a single Tb$^{3+}$ icosahedron and enumerate the lowest energy moment configurations.  We then calculate the  diffuse scattering from these configurations and compare with the experimental magnetic diffuse scattering measurements to identify the most likely single cluster moment configurations and find reasonable agreement between the broad features observed in our scattering simulations.  We further consider the role of higher-order (longer-range) intercluster correlations for the magnetic scattering.

\end{abstract}

\maketitle

\section{Introduction}
Quasicrystals are metallic compounds that manifest aperiodic, rather than periodic, long-range positional order and rotational symmetries (e.g. 5-fold, 10-fold) forbidden for conventional periodic crystals.\cite{Shechtman_1984}   Over the past 40 years, tremendous progress has been made in our understanding of the structure of quasicrystalline materials. In particular, the discovery of stable binary icosahedral quasicrystals \cite{Tsai_2000} and the elucidation of their structure via six-dimensional (6D) refinements have provided a detailed structural model of the icosahedral (\textit{i}) phase.\cite{Takakura_2007, Yamada_2016}  These studies have been greatly facilitated by the so-called crystalline approximants; large unit cell periodic crystals that feature local atomic motifs also found in the corresponding quasicrystalline phases.\cite{Goldman_1993}

The \textit{i}-\textit{R}-Cd (\textit{R} = Gd – Tm, Y)  family of magnetic quasicrystals\cite{Goldman_2013} and the closely related RCd$_{6}$ cubic approximants\cite{Gomez_2003} comprise an ideal set of model systems for attaining a deeper understanding of magnetism in quasicrystals since binary quasicrystals provide the compositionally simplest systems for the study of magnetic interactions in aperiodic compounds.  The \textit{R}Cd$_{6}$ approximants may be described, at ambient temperature, as a body-centered cubic packing of interpenetrating rhombic triacontahedra, or the so-called Tsai-type clusters.\cite{Tsai_2000} These clusters are comprised of four successive shells surrounding a tetrahedron of four Cd atoms: a dodecahedron composed of 20 Cd atoms; an icosahedron of 12 \textit{R} atoms, an icosidodecahedron of 30 Cd atoms; and an outermost shell described as a defect rhombic triacontahedron of 60 Cd atoms. These clusters are linked along the cubic axes by sharing a face and interpenetrating neighboring clusters along the body diagonal.\cite{Gomez_2003} Recent 6D structural refinements\cite{Yamada_2016} of \textit{i}-\textit{R}-Cd have shown that these Tsai-type clusters also form the backbone of the icosahedral phase, containing approximately 70\% of the Tb atoms, with the remainder located within the Friauf polyhedra in the “glue” that fills the gaps between the clusters.  These studies also suggest that up to 20\%\ of the \textit{R} ions on the icosahedron itself may be substituted by Cd, perhaps contributing to disorder related to magnetic frustration.

As the temperature is lowered, many RCd$_{6}$ approximants undergo a structural transition that is triggered by the ordering of the Cd tetrahedra at the center of each cluster resulting in a monoclinic unit cell.  For TbCd$_6$, this phase transition occurs at $T \approx 180$K, and the monoclinic unit cell, with space-group C2/c, is described as a $\sqrt{2}a \times a \times \sqrt{2}a$ superstructure of the high-temperature body-centered cubic unit cell.\cite{Nishimoto_2013}  As the temperature is further lowered, the \textit{R}Cd$_6$ (\textit{R} = Gd – Tm) magnetic approximants undergo a transition to an antiferromagnetic (AFM) ordered state at temperatures below T$_{N}$, ranging from approximately 23 K for \textit{R} = Tb to around 2 K for \textit{R} = Tm.\cite{Tamura_2010, Kim_2012, Mori_2012}  Preliminary analysis of neutron diffraction data from the TbCd$_6$ approximant phase indicate that the antiferromagnetic ordering between the Tb$^{3+}$ moments on each cluster follow the orientation of the inner Cd tetrahedra.  However, details of the arrangement of the moments on each cluster are, as yet, unresolved.  

In contrast to the magnetic order found for the some quasicrystal approximant phases, virtually all known magnetic quasicrystals manifest spin-glass-like behavior at low temperature.\cite{Hippert_2008, Sato_2005, Goldman_2014, Fisher_1999, Sebastian_2004} For the \textit{i}-\textit{R}-Cd icosahedral phases this occurs below the freezing temperature, T$ _{F} $, which varies from approximately 5.3 K for \textit{R} = Tb down to 2 K for \textit{R} = Tm.\cite{Kong_2014} Ferromagnetic order has been recently reported in icosahedral quasicrystals of \textit{i}-Au-Ga-Gd and \textit{i}-Au-Ga-Tb,\cite{Tamura_2021} and \textit{i}-Au-Si-Tb.\cite{Hiroto_2020} However, the more complex case of long-range antiferromagnetic order has not been found for icosahedral quasicrystals even though many theoretical studies suggest the possibility.\cite{Godreche_1986, Okabe_1988, Lifshitz_1999, Lifshitz_2000, Wessel_2003, Vedmenko_2004, Matsuo_2004, Jagannathan_2004, Thiem_2015} 
	
	Here, we describe elastic and inelastic neutron scattering measurements of the diffuse scattering from the icosahedral quasicrystal \textit{i}-Tb-Cd. We present a model for the arrangement of Tb$^{3+}$ moments on a single icosahedral cluster based on the analysis of the Crystalline Electric Field (CEF) level scheme determined from inelastic scattering measured using the Cold Neutron Chopper Spectrometer (CNCS) at the Spallation Neutron Source (SNS). Our calculation of the scattering from this model captures the broad features of the measured elastic diffuse scattering measured on CNCS and the Elastic Diffuse Scattering Spectrometer (CORELLI) at the SNS, and provides insight into the frustration that inhibits long-range magnetic order in the \textit{i}-\textit{R}-Cd quasicrystals.  However, we also see finer-scale structure in the elastic diffuse scattering measurements that indicates that magnetic correlations extend beyond a single cluster in the \textit{i}-phase. We attempt to gain some insight into the nature and role of these extended magnetic correlations through: (a) a simple model of packing icosahedral clusters (the "icosahedral glass" model\cite{Stephens_1986}) and (b) Monte Carlo simulations of spin correlations on and between neighboring icosahedra in the quasicrystalline structure obtained from the projection of the 6D structure of $i$-Tb-Cd determined previously.\cite{Yamada_2016} Both approaches yield finer scale structure which more closely resembles the measured magnetic neutron scattering patterns, but clearly additional work is required.  The Monte Carlo results provide further insight into the nature of magnetic frustration in these quasicrystalline systems.

\section{EXPERIMENTAL DETAILS}

The \textit{i}-Tb-Cd single-grain quasicrystals used for our neutron measurements were flux-grown\cite{Canfield_1992,Goldman_2013,Kong_2014,Canfield_2020} using isotopic $^{114}$Cd to minimize neutron absorption. The starting elements were packed in a frit-disc crucible set,\cite{Canfield_2020} with a molar ratio Tb:Cd = 0.8:99.2, which were then sealed in a silica tube under partial Ar atmosphere. The ampule was heated to 700$^{\circ}$C to realize a homogeneous liquid and slowly cooled to 330$^{\circ}$C. The remaining Cd solution was separated from the \textit{i}-Tb-Cd single grains in a centrifuge. Given the cost of $^{114}$Cd as well as the exceptionally limited range of the liquidus line for \textit{i}-\textit{R}-Cd in the binary \textit{R}-Cd phase diagrams, the nearly pure $^{114}$Cd decant was reused several times by adding in more Tb and regrowing i-Tb-Cd.\cite{Canfield_2016}  After several regrowths, the remaining decanted liquid was used to grow large single crystals of the crystalline approximant TbCd$_6$. Several of the larger single grains of the quasicrystal were co-oriented for the elastic and inelastic diffuse scattering measurements. A single crystal of the approximant was used for the inelastic scattering study.  

All neutron scattering measurements were performed at the SNS and at the High Flux Isotope Reactor (HFIR), both situated at Oak Ridge National Laboratory. The polarized neutron scattering measurements were performed on the HB-1 triple-axis spectrometer at HFIR. A co-aligned set of seven quasicrystals with a combined mass of $\approx$ 1 g was placed in a He-cryostat to reach the temperature range, 1.8 K $\leq$ T $\leq$ 30 K. The samples were aligned such that the two-fold plane was coincident with the horizontal scattering plane. Measurements were taken in the P $\parallel$ Q geometry (where P is the polarization vector and Q is the scattering vector), with an incident neutron energy of E$_{i}$ = 13.5 meV and collimations of 48' – 80' – sample – 80' – open. Pyrolytic graphite filters were placed both before and after the sample to reduce higher-order contamination. Heusler crystals were used as the monochromator and analyzer for the polarization setup with a flipping ratio of 10. 

Elastic and inelastic neutron scattering measurements were performed at the CNCS at the SNS,\cite{Ehlers_2011, Ehlers_2016} with an incident energy of E$_{i}$ = 10.04 meV using the same co-aligned set of samples. A background measurement taken at T = 30 K, above the spin-freezing temperature, was subtracted from the base temperature measurement at T = 2 K to extract the elastic magnetic diffuse scattering. The data were symmetrized and corrected for detector efficiency.  A single crystal of mass 2.7 g was used for the TbCd$ _{6} $ measurement at CNCS using the same configuration as that for the quasicrystal.

Elastic diffuse scattering measurements were performed at CORELLI\cite{Ye_2018} at SNS, which is a statistical chopper spectrometer with energy discrimination. A single quasicrystal of mass $\approx$ 200 mg, attached to the sample stick, was inserted in a He-cryostat for measurements at 2 K and 30 K. The sample was first aligned to have the horizontal scattering plane in the two-fold geometry, followed by remounting and realigning to measure the five-fold scattering plane. 

\section{RESULTS AND DISCUSSION}

\subsection{Polarized neutron scattering measurements}

\begin{figure}
  \centering
  \includegraphics[width=.8\linewidth]{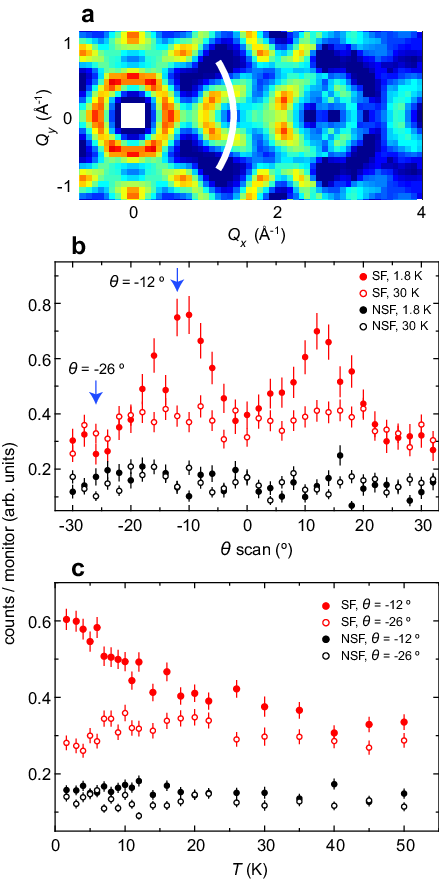}
  \caption{\label{Fig1}
           (Color online) Polarized neutron scattering measurements on \textit{i}-Tb-Cd accomplished on HB-1. The white arc in panel a shows the region of diffuse scattering, measured using unpolarized neutrons on CNCS, probed in the polarized beam measurements on HB-1.  Panel b displays the data taken in the spin-flip (SF, red) magnetic channel and non-spin-flip (NSF, black) nuclear scattering channel at T = 1.8 K (filled circles) and 30 K (open circles), respectively.  Panel c shows the temperature dependence of the magnetic and nuclear scattering, measured at the positions denoted by the blue arrows in panel b, demonstrating the onset of strong local magnetic correlations below approximately 30 K.}
\end{figure}

Our measurements, detailed below, demonstrate the presence of significant magnetic diffuse scattering from \textit{i}-Tb-Cd with overall icosahedral symmetry signaling the presence of strong short-range magnetic correlations at low temperature, similar to that found previously for R-based ternary quasicrystals such as \textit{i}-\textit{R}-Mg-Zn and \textit{i}-\textit{R}-Mg-Cd.\cite{Sato_2005} A small region of the unpolarized-beam elastic-diffuse neutron scattering, measured in the two-fold plane of i-Tb-Cd using CNCS, is shown in Fig. \ref{Fig1}a.  

The magnetic origin of the highly structured elastic diffuse scattering was confirmed by scans through the pattern using the polarized triple-axis spectrometer, HB-1, at the High Flux Isotope Reactor shown in Fig. \ref{Fig1}b.  Here, spin-flip scattering (SF) and non-spin-flip (NSF) scattering denote the magnetic and nuclear scattering contributions, respectively.  The magnetic data taken at 1.8 K, below the spin-freezing temperature of $\approx$5 K, demonstrates the emergence of structure associated with short-range magnetic correlations, whereas the nuclear scattering is constant.  

The temperature dependence of the magnetic and nuclear scattering is shown in Fig. \ref{Fig1}c, where we see that the structured magnetic diffuse neutron scattering fades with increasing temperature and is all but absent above approximately 30 K. A rough estimate of the magnetic correlation length, made from the measured widths of the diffuse scattering features, is 10 – 20 Å, similar to what was previously found for \textit{i}-Ho-Mg-Zn,\cite{Sato_1998} and consistent with the size of an icosahedron of Tb$^{3+}$ moments ($\approx$11 Å).  This observation motivated further study of the possible magnetic correlations between the Tb$^{3+}$ ions on a single icosahedron.

\subsection{Inelastic magnetic neutron scattering and the Tb$^{3+}$ single ion anistotropy}

\begin{figure}
  \centering
  \includegraphics[width=.8\linewidth]{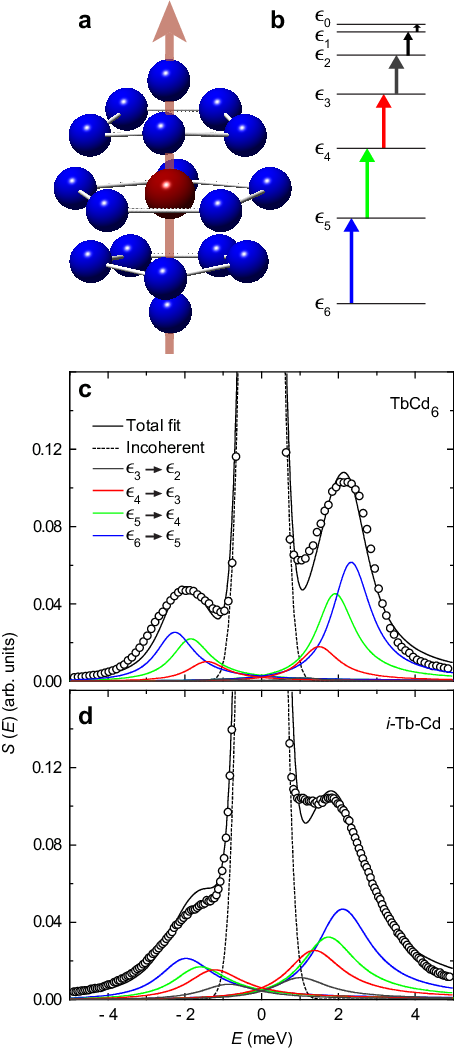}
  \caption{\label{Fig2}
           (Color online) Crystalline Electric Field (CEF) fits to the inelastic neutron scattering measured using the CNCS spectrometer. Panel a illustrates the local environment of a Tb$^{3+}$ ion (red) surrounded by Cd (blue) in a capped double pentagonal antiprism.  The unique axis is denoted by the arrow. Panel b shows a schematic of the CEF level scheme consisting of six doublets, $\epsilon_1$ through $\epsilon_6$, and one singlet, $\epsilon_0$.  The arrows denote transitions between the different CEF levels color-coded to the fit components in panels c and d.}
\end{figure}

The number of possible magnetic configurations for 12 Tb$^{3+}$ ions on a single icosahedron is essentially infinite.  However, information regarding the single ion anisotropy, via inelastic neutron scattering, can provide important information that significantly limits the number of magnetic configurations to be considered. Previous inelastic neutron scattering measurements on the TbCd$_6$ crystalline approximant found excitations below an energy transfer of 3 meV due to transitions among a manifold of crystalline electric field (CEF) levels.\cite{Das_2014} These excitations, and their evolution with temperature, were well characterized by considering only the leading term, B$_2^0O_2^0$, of the CEF level scheme (Fig. \ref{Fig2}b) for local pentagonal symmetry for the rare-earth ions (Fig. \ref{Fig2}a), indicating that the Tb$^{3+}$ moment is directed primarily along the unique local pseudo-five-fold axes of the Tsai-type clusters. We note here that this configuration corresponds to the magnetic texture, for $\theta$ = 0, considered by Watanabe.\cite{Watanabe_2021} Hiroto \textit{et al.} have also considered the CEF scheme for Tb$^{3+}$ in the quasicrystal approximant Au$_{70}$Si$_{17}$Tb$_{13}$ and, for this alloy, found that the Tb$^{3+}$ moments lie roughly in the plane perpendicular to the local pseudo-five-fold axes of the Tsai-type clusters.\cite{Hiroto_2020}  The crystal field excitations in Au$_{70}$Si$_{17}$Tb$_{13}$ are much broader than those we measured for TbCd$_6$, attributed to the chemical disorder inherent to the ternary approximant, and peak at higher energy ($\hbar\omega\approx 4$ meV).
	
The inelastic neutron scattering measured on CNCS for single crystals of both the crystalline approximant (TbCd$_6$) and the icosahedral phase (\textit{i}-Tb-Cd) are shown in Figs. \ref{Fig2}c and d.  These data were taken at T = 30 K for the samples oriented with their respective two-fold planes coincident with the scattering plane, and averaged over a range of momentum transfers (Q) at low scattering angles to increase counting statistics. The lines through the data represent fits using the CEF level scheme of Reference \onlinecite{Das_2014} allowing peak widths and energies to vary freely and the CEF parameters derived from the fits are given in Table \ref{Tab1}.  

The derived CEF parameters for the crystalline approximant are in good agreement with the previous powder measurement on the approximant phase, taken on IN4 at the Institut Laue-Langevin.  For the icosahedral phase, the CEF parameters reflect small variations from the values for the approximant and additional peak broadening is in evidence, likely due to variations in the local environment of the Tb$^{3+}$ ions.  Nevertheless, the agreement of the spectra and fitting parameters indicate that the CEF level schemes are quite similar and, therefore, we take the Tb$^{3+}$ moments to be directed primarily along the unique local five-fold axes of the Tsai-type clusters in the icosahedral phase as well.

\begin{table}
\caption{\label{Tab1}Results of the inelastic neutron scattering measurements on TbCd$_6$ and \textit{i}-Tb-Cd. The CEF parameter B$_2^0O_2^0$, energy of the transition to the first excited state, and full-width-at-half-maximum (FHWM) of the excitations for TbCd$_6$ measured at the IN4 time-of-flight spectrometer at the Institut Laue-Langevin, TbCd$_6$ measured at CNCS at the Spallation Neutron Source, and icosahedral \textit{i}-Tb-Cd measured at CNCS.\cite{Das_2014}}
\bigskip
\bigskip
\centering
\begin{tabular}{|c|c|c|c|}
\hline 
Sample & B$_2^0O_2^0$ & E(6$\rightarrow$5) (meV) & FWHM (meV) \\ 
\hline
TbCd$_6$ (IN4, ILL) & -0.073(1) & 2.41(4) & 1.0(2) \\ 
\hline 
TbCd$_6$ (CNCS, SNS) & -0.070(3) & 2.3(1) & 1.2(2) \\ 
\hline 
\textit{i}-Tb-Cd (CNCS, SNS) & -0.062(2) & 2.05(5) & 1.7(2) \\ 
\hline 
\end{tabular}
\end{table}

\subsection{Modeling moment configurations on the Tb$^{3+}$ icosahedral clusters}

The inelastic neutron scattering results allow us to consider a simplified model for the spin configurations on a single icosahedron of Tb$^{3+}$ moments for comparison with the results of our diffuse scattering measurements, detailed below.  Specifically, we consider that each Tb$^{3+}$ ion at the vertex of an icosahedron can assume only two directions; in, towards the center of the icosahedron along the local five-fold axis, or out.  This constraint admits only 2$^{12}$ = 4096 spin arrangements, considerably simplifying calculations of the relative energy for each possible moment configuration.  We calculated the energy of each Tb$^{3+}$ icosahedron moment arrangement from a magnetic Hamiltonian for classical spins as:
\begin{equation}
\label{eq1}
\begin{aligned}
H = -\sum_{i,j}J_{ij} \mathbf{S}_i\cdot\mathbf{S}_j 
\end{aligned}
\end{equation}
where $\mathbf{S}_i$ and $\mathbf{S}_j$ are moment vectors that point in or out along the local five-fold axis at each vertex, \textit{i} and \textit{j}. We consider only nearest-neighbor ($J_1$) and next-nearest-neighbor ($J_2$) interactions, so that $J_{ij}$ = $J_1$ or $J_2$, with $J$ positive(negative) for ferromagnetic(antiferromagnetic) interactions.

\begin{figure}
  \centering
  \includegraphics[width=1\linewidth]{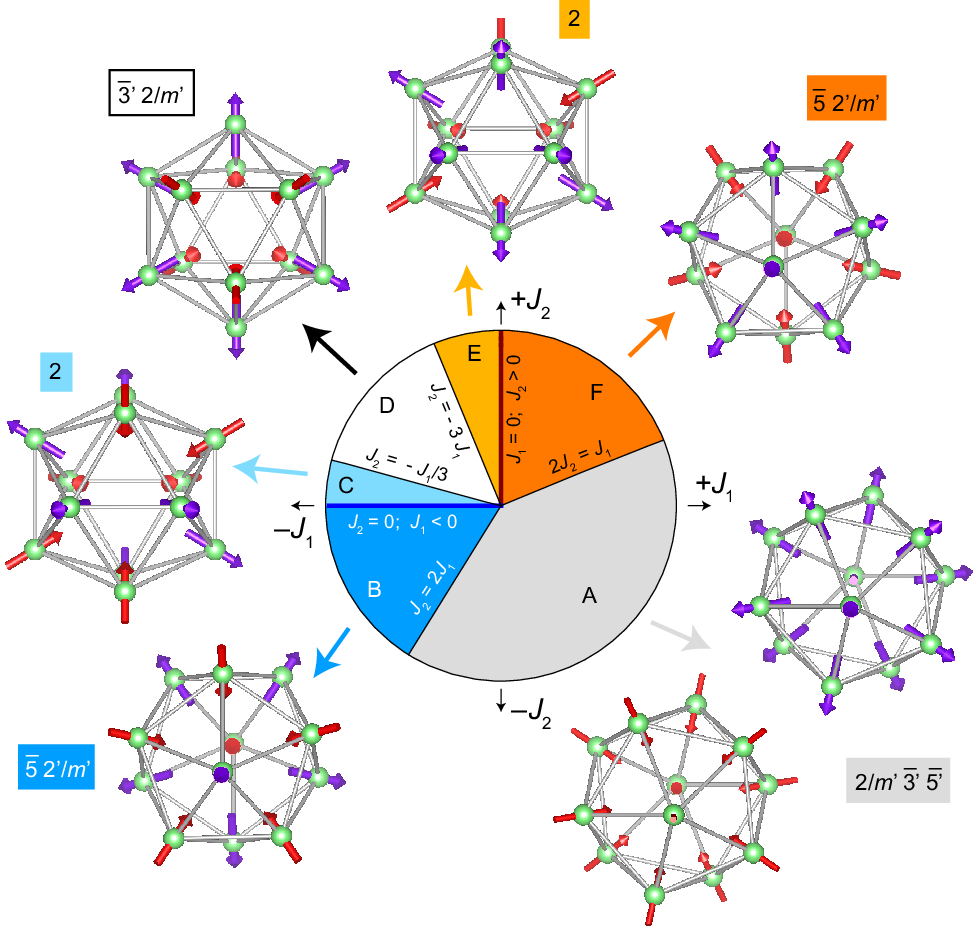}
  \caption{\label{Fig3}
           (Color online) Diagram of lowest energy moment configurations of the Tb$^{3+}$ icosahedra for ratios of $\frac{J_2}{J_1}$ along with the point symmetries preserved by the resulting moment configuration.  Moments directed radially out from the center of each icosahedron are denoted by purple arrows, and those directed inward are denoted by red arrows.    The net icosahedra moment directions in regions C and E are along the two-fold axes, for clusters in regions B and F along the five-fold axes.  The net icosahedra moments for regions A and D are zero.}
\end{figure}

	The lowest energy configurations for ratios of  $\frac{J_2}{J_1}$  are shown in Fig. \ref{Fig3} along with the point symmetries preserved by the resulting moment configuration.  These configurations, of course, feature a high degeneracy.  The degeneracy of the structures for each region, from A through F, is 2, 12, 60, 20, 60 and 12, respectively. The 60 clusters found in region C in Fig. \ref{Fig3} correspond to symmetry equivalent moment arrangements with a net cluster moment directed along one of the two-fold directions of the Tb$^{3+}$ icosahedron.  Similarly, the 12 degenerate moment configurations within region B correspond to symmetry equivalent moment arrangements with a net cluster moment directed along one of the five-fold directions of the Tb$^{3+}$ icosahedron.  For both regions, the the triangular faces features a (2 in/1 out) or (2 out/1 in) moment configuration. Interestingly, the exchange of only a single pair of Tb$^{3+}$ moment directions on the icosahedron is required to change the moment configurations from region B to C or vice-versa (this is also true for regions E and F in Fig. \ref{Fig3}). Region D corresponds to a net moment along one of the three-fold axes, and region A corresponds to a "hedgehog" or "unhappy-hedgehog" arrangement with no net moment for the cluster. The question remains as to which of these six lowest energy Tb$^{3+}$ icosahedron moment arrangements most adequately describes the local magnetic correlations found in \textit{i}-Tb-Cd.
	
\subsection{Comparison of elastic diffuse scattering measurements with scattering simulations}

We now turn to our measurements of the diffuse magnetic neutron scattering accomplished on the CNCS and CORELLI instruments at the SNS. In order to compare the measured patterns with the results of our analysis of moment configurations on a single icosahedron of Tb$^{3+}$ moments, the scattering intensity was calculated for each ground-state moment configuration as:
\begin{equation}
I(\mathbf{Q}) = \Bigl\vert f(Q) \sum_{i}\Bigl( \mathbf{S}_i - \frac{\mathbf{Q}\cdot\mathbf{S}_i}{Q}\Bigr) \exp{(i\mathbf{Q}\cdot\mathbf{R}_i)} \Bigr\vert^2 
\label{eq2}
\end{equation}			
where $\mathbf{S}_i$ is the $i^{th}$ Tb$^{3+}$ moment, $\mathbf{R}_i$ is the position of the $i^{th}$ moment with respect to the center of the icosahedron, and $f(Q)$ is the magnetic form factor for Tb$^{3+}$. The simulated scattering patterns were averaged over all symmetry equivalent moment arrangements within each region to yield the single-cluster diffuse scattering patterns for comparison with the measured data.    

\begin{figure}
  \centering
  \includegraphics[width=1\linewidth]{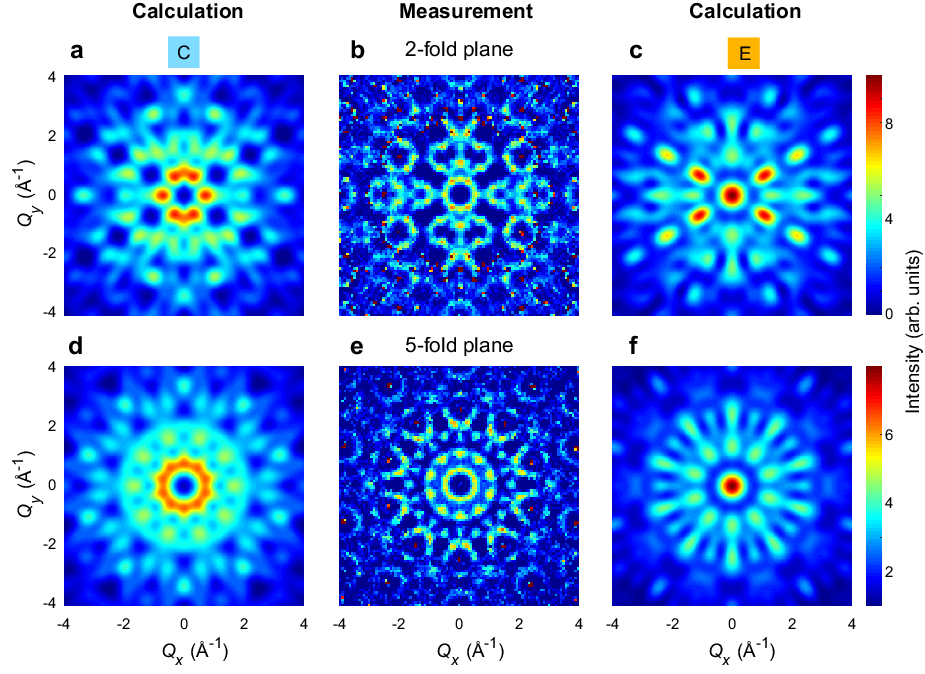}
  \caption{\label{Fig4}
           (Color online) Comparison of the experimental and simulated magnetic diffuse neutron scattering patterns measured for the two-fold and five-fold planes using CORELLI, calculated for symmetry averaged Tb$^{3+}$ icosahedron moment configurations corresponding to regions C and E in Figure \ref{Fig3}.}
\end{figure}

For comparison with the scattering calculations, the measured data taken at 30 K (where the structured diffuse scattering is absent) was subtracted from the data taken at 2 K (below the spin-freezing temperature) and the patterns were symmetrized. For example, two sets of the calculated diffuse scattering patterns are shown in Fig. \ref{Fig4}, corresponding to averages over the 60 configurations found in regions C and the 60 configurations of Region E of Fig. \ref{Fig3}. The scattering patterns for region C, corresponding to dominant AFM $J_1$, produce features of the correct scale and placement relative to the measured pattern, whereas that for region E (dominant FM $J_2$) does not.     Comparisons of the diffuse scattering patterns for all configurations within regions A - F are shown in the Appendix.

\begin{figure}
  \centering
  \includegraphics[width=1\linewidth]{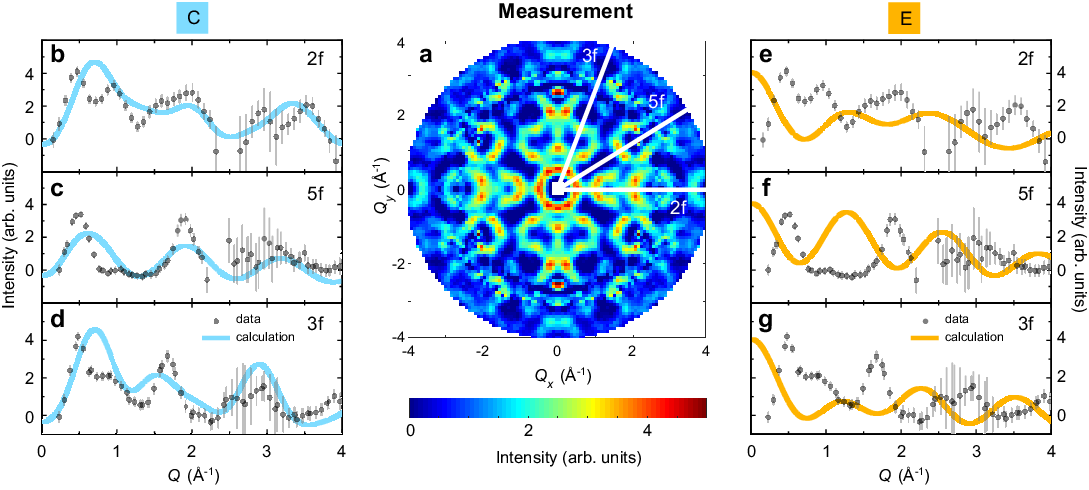}
  \caption{\label{Fig5}
           (Color online) Comparison of line scans through the magnetic diffuse neutron scattering for the Tb$^{3+}$ icosahedron moment configurations corresponding to regions C and E in Figure \ref{Fig3}. The white lines in panel a denote the high-symmetry two-fold (2f), three-fold (3f) and five-fold (5f) axes of the icosahedral structure.  The data, measured at CNCS,  are shown as points with error bars in panels b through g and the solid lines denote the cuts through the calculated diffuse scattering patterns along these same directions.}
\end{figure}

	Figure \ref{Fig5} provides further detail, showing cuts through the calculated scattering patterns along the high-symmetry two-, three- and five-fold axes in the two-fold plane for regions C and E, and the elastic diffuse neutron scattering measured on CNCS. Again, the cuts through the calculated pattern for region C, multiplied by a single scale factor in Figs. \ref{Fig5}b - d capture the overall scale and modulation of the scattering of the measured data, whereas the cuts extracted from region E (Figs \ref{Fig5}e – g) are clearly at odds with the measurement.  However, there is also additional fine-scale structure in the data shown in both Figs. \ref{Fig4} and \ref{Fig5} that is not well reproduced by the calculation. These differences most likely arise from intercluster magnetic correlations that are not captured by a single-cluster model as discussed below.  
	
Inspection of all of the simulated elastic diffuse scattering patterns reveals that the closest agreement with the experimental data is found for the averaged moment arrangement in region C. However, as shown in the Appendix, the 12 moment arrangements within region B are also in reasonable agreement with the measured magnetic diffuse scattering along the two- and three-fold axes. Given the close similarity between the two arrangements (requiring the exchange of moment directions for only a single pair of Tb$^{3+}$), the configurations within both regions B and C likely contribute to the scattering.  Regardless, these comparisons demonstrate that the dominant magnetic interaction is antiferromagnetic $J_1$ between nearest-neighbor moments on the Tb$^{3+}$ icosahedron, consistent with the negative Curie-Weiss temperature found in bulk measurements of the magnetic susceptibility for all \textit{i}-R-Cd quasicrystals.\cite{Kong_2014} The antiferromagnetic interactions between moments arranged on the triangular faces of the icosahedron naturally give rise to geometrical frustration within the fundamental magnetic clusters of the icosahedral phase.

\subsection{Heuristic modeling of intercluster correlations}

Although the diffuse scattering for magnetic configuration in region C of Fig. \ref{Fig3} reproduces the broad-scale features of the measured magnetic diffuse scattering, there is additional finer-scale structure in Figures \ref{Fig4} and \ref{Fig5} that is not well reproduced by the single-icosahedron scattering calculation.  Here, we present a very rudimentary model of inter-icosahedron correlations to illustrate one approach to incorporating longer length scale magnetic interactions. 
This approach is rooted in previous work employing the icosahedral glass model\cite{Stephens_1986} to describe nuclear diffuse scattering in some icosahedral alloys.\cite{Goldman_1988} It is also worth mentioning here that the $i$-Tb-Cd quasicrystals do exhibit disorder characterized by a moderate degree of frozen-in phason strain\cite{deboissieu_2011}, endemic to many icosahedral quasicrystals.  This phason strain corresponds to errors or defects in the packing of structural subunits, similar to the more extreme case of an icosahedral glass. In both this section, and the next, we assume full occupancy of Tb$^{3+}$ on the icosahedra.

\begin{figure}
  \centering
  \includegraphics[width=.8\linewidth]{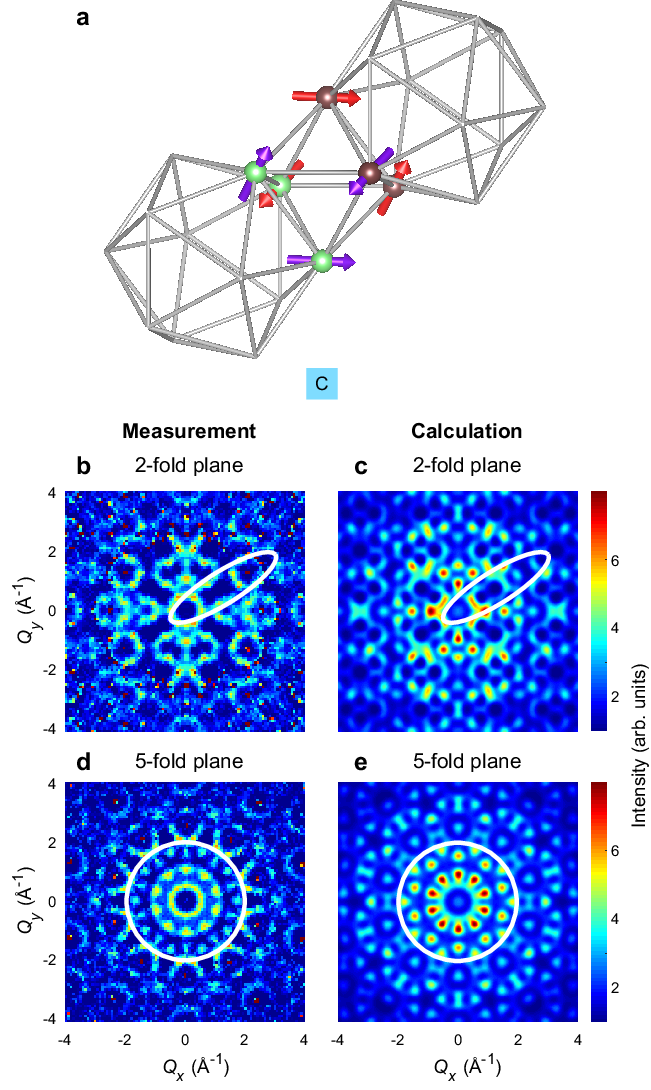}
  \caption{\label{Fig6}
           (Color online) Two-icosahedra diffuse scattering.  Panel a shows an example of a connection between two Tb$^{3+}$ icosahedra of region C in Fig. \ref{Fig3}.  The moments at the boundary are denoted by arrows pointing in or out from their respective icosahedra. These six Tb$^{3+}$ ions form a nearly perfect octahedron and represent the lowest energy moment configurations on the octahedra that connect adjacent Tb$^{3+}$ moments along three-fold directions. Panels b and c display the measured magnetic diffuse neutron scattering in the two-fold plane at CORELLI and the scattering calculated from the model described in the text.  Panels d and e display the measured magnetic diffuse scattering in the five-fold plane at CORELLI and the scattering calculated from the model described in the text.}
\end{figure}

Each icosahedral arrangement of Tb$^{3+}$ moments features 20 triangular faces normal to the three-fold directions.  If we connect two of these icosahedral with their faces rotated by 60 degrees, the Tb$^{3+}$ ions on the boundaries of adjacent icosahedra form a nearly perfect octahedron since the inter-icosahedron Tb–Tb distance is nearly the same as the intra-icosahedron Tb–Tb separation, as shown in Fig. \ref{Fig6}a.  In TbCd$_6$ each Tsai-type cluster is coordinated by eight other clusters joined in this manner along the three-fold directions of the body-centered cubic lattice.  However, for the icosahedral phase, the coordination number is variable with the most frequently encountered coordination number of approximately 5.5.\cite{Takakura_2008,Takakura_2017} 

	Our model considers that two icosahedra can be connected along any of the three-fold directions with equal probability (1/20) with the arrangement of neighboring icosahedra obeying steric constraints. That is, two icosahedra attached to a third do not interfere with each other. The diffuse scattering is calculated from an average over the scattering due to the 20 possible two-icosahedra configurations:	
	
\begin{equation}
\label{eq3}
\begin{aligned}
I_{\text{two-icos}}(\mathbf{Q}) = \bigl\langle \left\vert 1+P_k\exp{(i\mathbf{Q}\cdot\mathbf{R}_k)} \right\vert^2 \bigr\rangle
\end{aligned}
\end{equation}	
where the $\mathbf{R}_k$ are the twenty inter-icosahedra vectors along three-fold axes and, as mentioned above, $P_k$ = $\dfrac{1}{20}$.  If we simply multiply the two-icosahedra structure factor by the single-icosahedron magnetic scattering for region C, shown in Figs. \ref{Fig4}a and d, the scattering patterns in Figs. \ref{Fig6}c and e result.  These results are compared with the experimental data from CORELLI in Fig. \ref{Fig6}b and d. 

We find that even this simple model for FM inter-icosahedra magnetic correlations provides additional fine scale detail observed in the regions outlined by the white ovals in Figs. \ref{Fig6}b and c and the circles in Figs. \ref{Fig6}d and e.  For example, in the fivefold pattern, the more intense scattering ring close to the center of the pattern in the single-cluster calculation is much reduced, and the 20 diffraction maxima in a corrugated ring near the boundary of the circle are now in evidence.  In the two-fold pattern, the structure of the voids in the scattering are more closely modeled.  However, the details of features in both the two-fold and five-fold planes are not fully captured in this simple treatment and, of course, a proper calculation of the “two-icosahedra structure factor” for the moment arrangements must take into account arrangements of moments from other regions in Fig. \ref{Fig3} (such as region B) which may come into play. Nevertheless, our inter-cluster modeling does point to a possible path forward in fully describing the diffuse magnetic neutron scattering for \textit{i}-Tb-Cd and other magnetic icosahedral \textit{i}-R-Cd quasicrystals.
	
Finally, we recall that only about 70\% of the Tb$^{3+}$ ions are associated with the icosahedron in the Tsai-type clusters of the quasicrystalline phase, with the balance located in the “glue” that fills the gaps between the clusters.  Our modelling of the magnetic diffuse scattering assumes that these Tb$^{3+}$ moments contribute only to the featureless paramagnetic background. 

\subsection{Monte-Carlo simulations of multi-cluster model}
\begin{figure}
\centering
\includegraphics[width=1\linewidth]{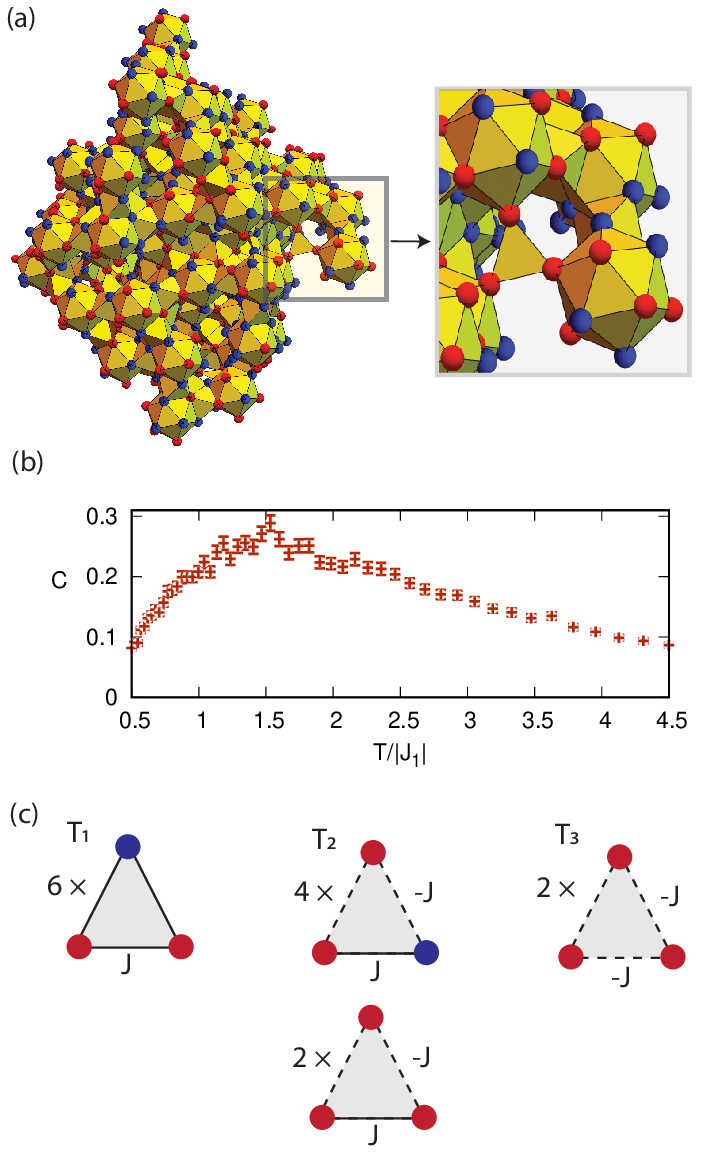}
\caption{(color online)(a) Illustration of the simulated quasicrystal model composed of $N_c = 72$ connected icosahedral clusters. Each cluster hosts $12$ Ising spins $S_i$ that are denoted by red and blue spheres, corresponding to $S_i = 1$ and $-1$, respectively. Nearest-neighbor bonds are shown in black (both for intra- and inter-cluster spin pairs), and the faces of all nearest-neighbor triangles are drawn in yellow. The distribution of red and blue spheres represents a thermalized snapshot spin configuration. In the center of the inset one finds a triangle, where the spins are located on three different clusters. This triangle is thus found in its ground state configuration (all vertices are red), since spin pairs on different clusters interact FM. Further to the right one observes a triangle that connects two clusters, which is in one of its red-blue-blue ground state configurations. (b) The specific heat as a function of temperature, $T$, exhibits a broad peak at $T = 1.5 |J_1|$ which is characteristic of a spin glass (c) Representative ground state spin configuration of the three classes of triangles $T_1$ (left), $T_2$ (center), and $T_3$ (right) that exist in the model. The Ising spins $S_i$ that are connected by solid lines are on the same cluster and thus interact AFM. In contrast, those connected by dashed lines reside on different clusters and thus interact FM in the effective Ising model. The numbers denote the ground state degeneracy of the triangle. The other ground states are explicitly stated in the text.}
\label{fig:theory_1}
\end{figure}

To further elucidate the role of inter-cluster correlations, we performed large-scale Monte Carlo (MC) simulations of a three-dimensional quasiperiodic spin system composed of $N_c=72$ connected icosahedral clusters. The spins are located on the vertices of the icosahedra such that the model contains $N=72 \times 12 = 864$ spins in total. The real-space locations of the spins $\boldsymbol{R}_i$ were obtained by projecting a higher-dimensional decorated model determined from single-grain diffraction measurements on $i$-Tb-Cd\cite{Yamada_2016} onto three dimensions [see Fig.~\ref{fig:theory_1} (a)] and choosing 72 clusters from the interior of the projection. 

Single-ion anisotropy forces each spin $\boldsymbol{S}_i$ to point along a local five-fold axis $\boldsymbol{n}_i$ and the spins $\boldsymbol{S}_i = S_i \boldsymbol{n}_i$ can thus be modeled by Ising degrees of freedom $S_i = \pm 1$. Based on our previous analysis of the single cluster model, we considered an antiferromagnetic (AFM) (Heisenberg) spin interaction $J_1 < 0$ between nearest-neighbor spin pairs, such that the Hamiltonian of Eqn. (\ref{eq1}) reads 
\begin{equation}
\label{eq:theory_ham}
\begin{aligned}
  H = - J_1 \sum_{\langle i, j\rangle }\boldsymbol{S}_i \cdot \boldsymbol{S}_j = - J_1 \sum_{\langle i, j \rangle} S_i S_j \, \boldsymbol{n}_i \cdot \boldsymbol{n}_j \\= -\sum_{\langle i, j \rangle} \tilde{J}_{1;ij} S_i S_j
\end{aligned}
\end{equation}
Here, we have defined an effective Ising interaction constant $\tilde{J}_{1;ij} = J_1 \boldsymbol{n}_i \cdot \boldsymbol{n}_j$. Nearest-neighbor spins can be located on the same or different clusters. These situations are illustrated in the inset of Fig.~\ref{fig:theory_1}(a). In both cases, the spins are separated by a real-space distance $d_1 \approx 5.79$~\AA~and they thus exhibit the same AFM interaction $J_1$. The different geometry, however, affects the sign of the effective Ising interaction constant $\tilde{J}_1$. Specifically, the scalar product of the two local five-fold axes $\boldsymbol{n}_i \cdot \boldsymbol{n}_j = \pm 0.45$ has equal magnitude but opposite sign depending on whether the spins $\boldsymbol{S}_i$ and $\boldsymbol{S}_j$ are on the same or neighboring clusters. If they are located on the same cluster, we find $\boldsymbol{n}_i \cdot \boldsymbol{n}_j > 0$, whereas $\boldsymbol{n}_i \cdot \boldsymbol{n}_j < 0$ if they reside on different clusters. As a result, Ising spins on the same cluster interact via an AFM effective Ising spin interaction $\tilde{J}_1 < 0$, but spins on different clusters interact ferromagnetically with $\tilde{J}_1 > 0$. In the following, we drop the tilde notation and use $J_1$ to refer to the Ising interaction strength in our effective Ising model. 

To simulate the finite temperature behavior of the quasicrystal Ising model, we performed parallel-tempering MC simulations, where each MC step consists of a local Metropolis update followed by a parallel-tempering move~\cite{Marinari_1992, Hukishima_1996}. Our simulations used $2.048 \times 10^{10}$ MC steps in total, the first half of which was discarded to allow for thermalization. After thermalization, the observables were measured every $10^{7}$ MC steps and we performed a total of $1024$ measurements. Errors were calculated using the standard Jackknife method~\cite{Efron_1982}. We simulated $50$ logarithmically spaced temperatures between $0.5 \leq T/J_1 \leq 4.5$ in parallel. In Fig.~\ref{fig:theory_1}(b) we show the specific heat versus temperature, $T$, exhibits a broad peak around $T = 1.5 |J_1|$, that is reminiscent of spin glass behavior across the freezing transition. 

Next, we studied the local arrangement of spins on the fundamental triangular units of the system, which yields insight into the type of frustration in the model. The system has $T_{\text{tot}} = 2546$ spin triangles in the system, which can be separated into three classes, shown in Fig.~\ref{fig:theory_1}(c). There are $T_1 = 1440$ triangles where all spins belong to the same icosahedron cluster, $T_2 = 834$ triangles where spins are shared between two clusters, and $T_3 = 272$ triangles where all three spins belong to different clusters. The effective Ising exchange interaction across intracluster bonds is antiferromagnetic, while it is ferromagnetic across intercluster bonds. Therefore, the triangle classes exhibit different energy spectra of the eight possible spin states $\ket{s_1 s_2 s_3}$ with $s_i \in \{1 \equiv \, \uparrow, -1 \equiv \, \downarrow\}$. Figure~\ref{fig:theory_1}(c) shows a representative ground state configuration for the respective triangle classes and their degeneracy. Specifically, the lowest energy states on the $T_1$ triangles are the (2-in/1-out) and (2-out/1-in) states $\{\ket{\downarrow \uparrow\uparrow}, \ket{\uparrow \downarrow \uparrow}, \ket{\uparrow\uparrow\downarrow}, \ket{\downarrow\downarrow\uparrow}, \ket{\downarrow\uparrow\downarrow}, \ket{\uparrow \downarrow\downarrow}\}$ with energy $E_0 = -J_1$, while the two excited states $\{\ket{\uparrow\uparrow\uparrow}, \ket{\downarrow\downarrow\downarrow}\}$ have energy $E_1 = 3J_1$. In contrast, the $T_2$ triangles connect two clusters. Taking the bonds between $(s_2, s_3)$ and between $(s_3, s_1)$ to be FM, the $T_2$ ground states are thus $\{\ket{\uparrow\uparrow\uparrow}, \ket{\downarrow\downarrow\downarrow}, \ket{\uparrow\downarrow\uparrow}, \ket{\uparrow\downarrow\downarrow}, \ket{\downarrow\uparrow\uparrow}, \ket{\downarrow\uparrow\downarrow}\}$ with energy $E_0 = -J_1$, while the two excited states are $\{\ket{\downarrow\downarrow\uparrow}, \ket{\uparrow\uparrow\downarrow}\}$ with energy $E_1 = 3J_1$. Finally, the lowest energy states on the $T_3$ triangles, where all spins interact FM, are $\{\ket{\uparrow\uparrow\uparrow}, \ket{\downarrow\downarrow\downarrow}\}$ with energy $E_0 = -3J_1$. The remaining six excited states have energy $E_1 = J_1$. 

\begin{figure}
\centering
\includegraphics[width=1\linewidth]{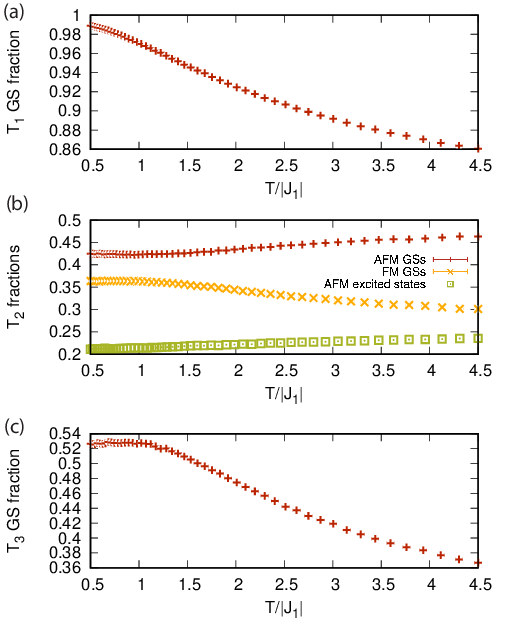}
\caption{(color online) Average fraction of $T_1$, $T_2$, and $T_3$ triangles (a-c) that are in their ground state configuration. Panel (b) distinguishes AFM and FM ground states [see Fig.~\ref{fig:theory_1}(c)] and also contains the fraction of triangles in excited states. As the temperature $T$ is lowered, the fraction of triangles in a ground state increases, but $20\%$ of $T_2$ and $50\%$ of $T_3$ triangles remain in an excited state down to the lowest temperatures. The results are obtained as an average of $1024$ measurements that occur every $10^{7}$ MC steps after an initial thermalization period of $10^{10}$ MC steps. Error bars are obtained using the Jackknife method. }
\label{fig:theory_2}
\end{figure}

In Fig.~\ref{fig:theory_2}, we show the average number of triangles of type $T_1$, $T_2$, and $T_3$, where the spins are in their lowest energy configuration. For the $T_2$ triangles that connect two clusters, we separate the four AFM ground state configurations from the two FM ones. We also graph the number of $T_2$ triangles with excited spin configurations. As the temperature is reduced, the number of triangles in a ground state configuration increases, and in Fig.~\ref{fig:theory_2}(a), we observe that almost all ($\approx 99\%$) of the $T_1$ triangles reach their ground state. These correspond to the $72$ ground states of a single icosahedron in phase B and C (see Fig.~\ref{Fig3}). However, the situation is different on the triangles that connect different clusters. As shown in Fig.~~\ref{fig:theory_2}(b) and (c), about $20\%$ of the $T_2$ triangles and $\approx 50\%$ of the $T_3$ triangles remain in an excited state configuration at low temperatures. This is typical for glassy systems that host a complex energy landscape with an large number of local minima. In the quasicrystal Ising model, the behavior results from the strong magnetic frustration that leads to a large number of degenerate spin configurations on a single-cluster. As shown in the previous section, there are $72$ degenerate single cluster states counting both states in phase B (12 states) and C (60 states). While these states are degenerate on a single disconnected cluster, different cluster configurations cannot be connected in a way that satisfies every inter-cluster bond and a large fraction of triangles $T_2$ and $T_3$ remains in an excited state. The inter-cluster connections impose ``longer-range'' constraints that break the degeneracy between different (isolated) cluster configurations, which leads to a complex energy landscape.

\begin{figure}
\centering
\includegraphics[width=1\linewidth]{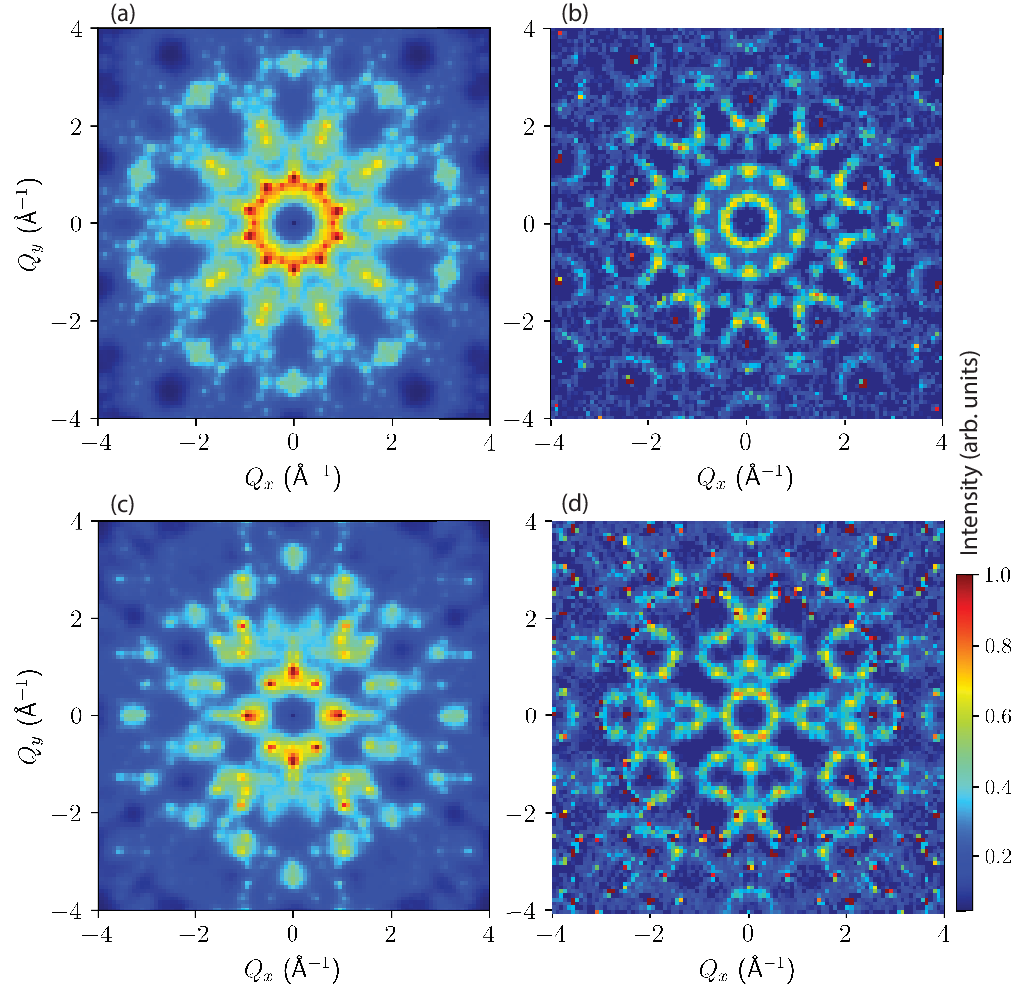}
\caption{(color online) (a, c) MC results of the structure factor $I(\boldsymbol{q})$ in the $N_c = 72$ quasicrystal Ising model at temperature $T = 1.5~J_1$. Panel (a) shows the result in the fivefold plane and (c) in the twofold plane. The results are obtained by averaging $1024$ MC snapshots that are taken every $10^7$ MC steps, after an initial thermalization period of $10^{10}$ steps. (b, d) Experimental neutron scattering results in the (b) fivefold and (d) twofold plane.  }
\label{fig:theory_3}
\end{figure}

\begin{figure}
\centering
\includegraphics[width=.8\linewidth]{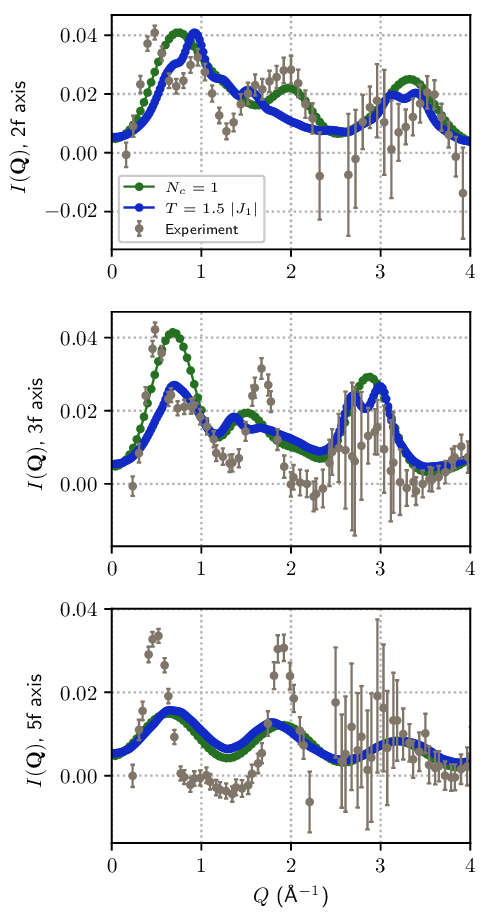}
\caption{(color online) (a) Cut along the twofold axis in the twofold plane. (b) Cut along the threefold axis. (c) Cut along the fivefold axis. Green dots are results for a single cluster, where we averaged over all $72$ ground states in phase B and C with equal weight. Blue dots are MC results for the $N_c = 72$ system at $T = 1.5|J_1|$ and correspond to cuts through Fig.~\ref{fig:theory_3}(c). The tan dots are neutron scattering results.}
\label{fig:theory_4}
\end{figure}

We have also calculated the averaged structure factor $I(\boldsymbol{q})$ at temperature $T = 1.5~|J_1|$ using Eq.~\eqref{eq2}. We chose this temperature since it corresponds to the location of the broad peak in the specific heat [see Fig.~\ref{fig:theory_1}(b)]. We obtained $I(\boldsymbol{q})$ by averaging over $1024$ statistically independent spin snapshot configurations obtained in the MC simulation. The snapshots were taken every $10^{7}$ MC steps after an initial period of $1.024 \times 10^{10}$ thermalization steps. Fig.~\ref{fig:theory_3} shows results in the fivefold (a) and twofold (b) scattering plane, together with the experimental data from Fig.~\ref{Fig4}. One-dimensional cuts through the data along the twofold, threefold, and fivefold directions are shown in Fig.~\ref{fig:theory_4}. Overall, we find reasonable agreement with the experimental result and can identify several common features. The simulation reproduces the characteristic ring of scattering and the regions of low intensity in the fivefold plane and the region of low scattering around the fivefold direction in the twofold plane. However, we can also identify features in the experimental data that are not captured in the model, such as the elongated regions of low intensity discussed in Fig.~\ref{Fig6}, which points towards the importance of further neighbor interactions and quenched disorder that is not included in the quasicrystalline model. The importance of disorder is further corroborated by the success of the heuristic quasicrystal glass model in capturing regions of low scattering intensity, as discussed in the previous section and Fig.~\ref{Fig6}. 

In Fig.~\ref{fig:theory_4}, we show one-dimensional cuts along high-symmetry directions in the twofold plane to make a more quantitative comparison to the experimental data and the single cluster model in phase C. Along the twofold direction [see Fig.~\ref{fig:theory_4}(a)], we observe that each of the broad single-cluster peaks at $\approx 0.8$~\AA$^{-1}$ and $\approx 3.2$~\AA$^{-1}$ split into two peaks in the $N_c = 72$ model. This peak splitting is an good agreement with the experimental data. The single cluster peak at $\approx 2.0$~\AA$^{-1}$ is shifted towards smaller momentum values in the $N_c = 72$ model, and the shifted location coincides well with an experimental peak. However, in contrast to the single cluster model, the quasicrystal model does not capture the experimental peak at $\approx 2.0$~\AA$^{-1}$. 
Along the threefold direction [see Fig.~\ref{fig:theory_4}(b)] we observe five peaks in the experimental data, but only four peaks in the simulation. The peaks in the quasicrystal model are slightly shifted compared to the experiment and the peak height found in the simulation also shows some disagreement with experiment. While the experiment shows a pronounced peak at $\approx 0.5$~\AA$^{-1}$ and a smaller one nearby at $\approx 0.9$~\AA$^{-1}$, the quasicrystal model exhibits a single peak at $\approx 0.8$~\AA$^{-1}$ with a peak maximum close to the smaller experimental peak height. Interestingly, the single cluster model exhibits a single peak at $\approx 0.7$~\AA$^{-1}$, whose height closely matches that of the dominant experimental one. While there exists a third experimental peak at $1.6$~\AA$^{-1}$, the model shows a peak at a slightly smaller wavevector $\approx 1.3$~\AA$^{-1}$. Finally, two experimental peaks at $2.5$~\AA$^{-1}$ and $3.0$~\AA$^{-1}$ are well captured by the quasicrystal model simulations. 
Along the fivefold direction [see Fig.~\ref{fig:theory_4}(c)], the quasicrystal results more closely agree with the single cluster model in phase C. Both show a slightly lower intensity compared to the other high-symmetry directions, which is in agreement with experiment. The location of these peaks is in reasonable agreement with experiment, even though the peak heights in the simulation are smaller by a factor of two.

\section{Summary}
Summarizing, we have used elastic and inelastic neutron scattering from single-grain isotopically-enriched samples to elucidate the local magnetic correlations between Tb$^{3+}$ moments in quasicrystalline \textit{i}-Tb-Cd.  The inelastic neutron scattering measurements of the CEF excitations demonstrated that the Tb$^{3+}$ moments are directed primarily along the local five-fold axes of the Tsai-type cluster as was found for the TbCd$_6$ approximant phase.\cite{Das_2014} This allowed the consideration of a simplified Ising-type model for the moment configurations, which we first examined for a single Tb$^{3+}$ icosahedron. A survey of the lowest energy moment configurations on a single icosahedron, as a function of first and second neighbor couplings, and the comparison with the elastic diffuse neutron scattering measurements indicates that the spins on each triangular face of the cluster feature a (2 in/1 out) or (2 out/1 in) moment configuration.  The antiferromagnetic interaction between moments arranged on the triangular faces of the icosahedron gives rise to geometrical frustration with a large ground state degeneracy. The best candidate spin structure(s) find reasonable agreement with the measured magnetic diffuse scattering although longer-range magnetic correlations are also evident.

In an attempt to elucidate the nature of these longer range correlations we investigated two models which assume full occupancy of the Tb icosahedra.  The first makes use of the icosahedral glass model to extend the structure beyond a single cluster albeit with significant intrinsic disorder.  Nevertheless, this approach yields diffuse scattering profiles in closer agreement with our measurements.  We also employed MC simulations of a quasicrystal model obtained by a projection of the 6D atomic structure down to three dimensions. The $N_c = 72$ clusters reproduces several general features of the experimental structure factor, which indicates a dominant, antiferromagnetic nearest-neighbor interaction in the material. On a single icosahedron, this leads to a large ground state degeneracy with $72$ degenerate spin configurations (phase B and C). Connecting clusters via nearest-neighbor bonds imposes additional constraints that cannot be simultaneously satisfied, leaving behind bonds, where spins remain in an excited state. This lifts the ground state degeneracy and leads to a complex energy landscape with a large number of local minima. We observe that $20\%$ of the triangles that connect two clusters and $50\%$ of those that connect three clusters, remain in an excited state at low temperatures. This is further corrobated by the behavior of the specific heat that exhibits a broad peak around $T \approx 1.5|J_1|$, consistent with spin glass behavior. Finally, the fact that some features in the experimental data are not captured in the model simulations indicates the importance of further neighbor spin-spin interactions and quenched disorder in the material. Taking these effects into account in a theoretical model simulation is an interesting direction for future study.
 
\begin{acknowledgments}
The authors would like to thank M. deBoissieu, T. J. Sato and R. Tamura for very useful discussions.
P.P.O. thanks A. Vishwanath and Harvard University for hospitality during the final stages of this project. This research used resources at the High Flux Isotope Reactor and Spallation Neutron Source, a DOE Office of Science User Facility operated by the Oak Ridge National Laboratory.  This work was supported by the U.S. Department of Energy, Office of Basic Energy Science, Division of Materials Sciences and Engineering. The research was performed at the Ames Laboratory. Ames Laboratory is operated for the U.S. Department of Energy by Iowa State University under Contract No. DE-AC02-07CH11358.

\end{acknowledgments}

\appendix
\section*{APPENDIX}
Figures \ref{FigA1} through \ref{FigA6} compare scans through the magnetic diffuse neutron scattering for Tb$^{3+}$ icosahedron moment configurations corresponding to regions A through F, respectively. Panel a of each Figure shows the scattering pattern measured for the two-fold plane using the CNCS spectrometer at the Spallation Neutron Source.  The white lines in panel a denote the high-symmetry two-fold (2f), three-fold (3f) and five-fold (5f) axes.  Panels b through d show cuts through the diffuse magnetic neutron scattering along these high-symmetry directions.  The points denote the measured data and the solid lines denote the calculated curves.   Panel e shows the calculated magnetic diffuse neutron scattering pattern in the two-fold plane averaged over the magnetic configurations for each region in Figure \ref{Fig3} as described in the main text.
\clearpage

\begin{widetext}

\begin{figure}
  \centering
  \includegraphics[width=1\linewidth]{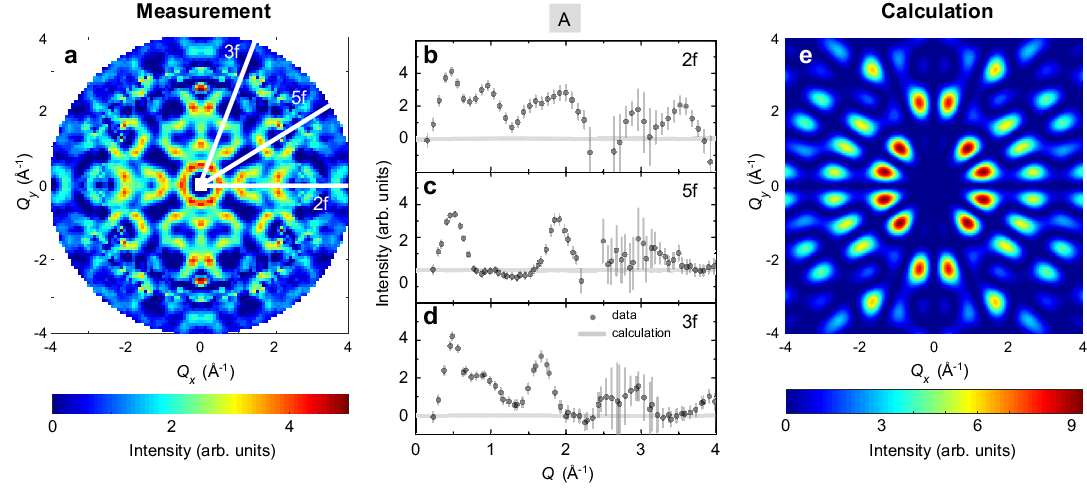}
  \caption{\label{FigA1}
           (Color online) Comparison of scans through the averaged magnetic diffuse scattering for Tb$^{3+}$ icosahedron moment configurations corresponding to region A of Figure \ref{Fig3}.}
\end{figure}

\begin{figure}
  \centering
  \includegraphics[width=1\linewidth]{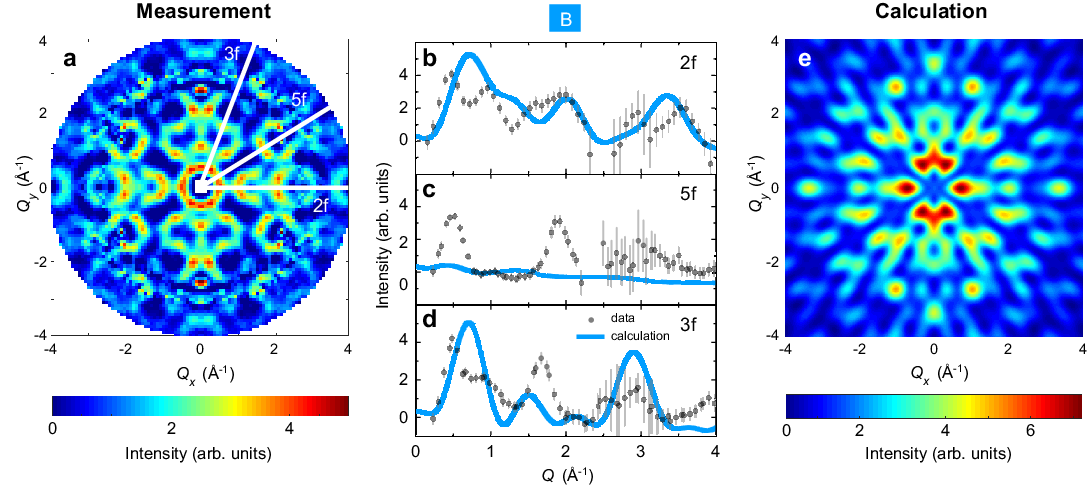}
  \caption{\label{FigA2}
           (Color online) Comparison of scans through the averaged magnetic diffuse scattering for Tb$^{3+}$ icosahedron moment configurations corresponding to region B of Figure \ref{Fig3}. }
\end{figure}
\clearpage
\clearpage

\begin{figure}
  \centering
  \includegraphics[width=1\linewidth]{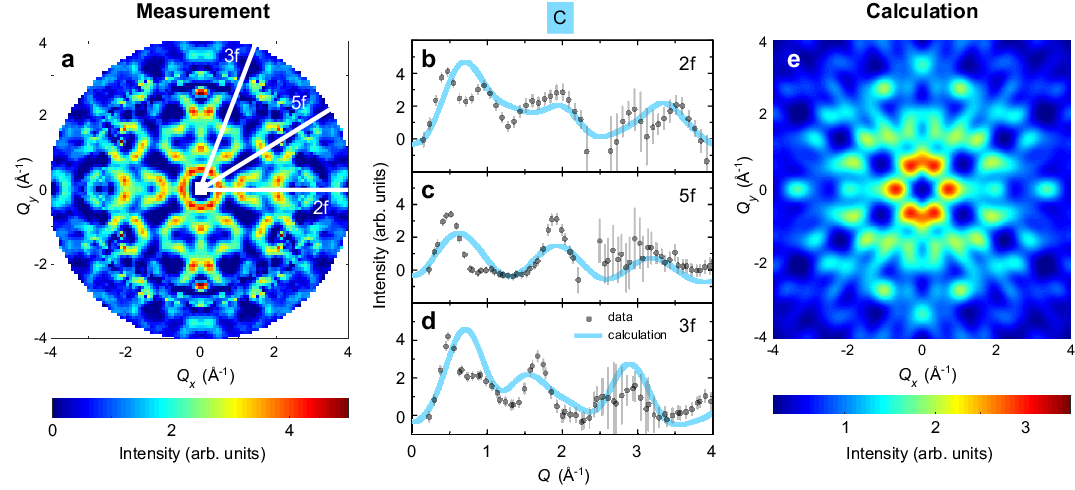}
  \caption{\label{FigA3}
           (Color online) Comparison of scans through the averaged magnetic diffuse scattering for Tb$^{3+}$ icosahedron moment configurations corresponding to region C of Figure \ref{Fig3}. }
\end{figure}

\begin{figure}
  \centering
  \includegraphics[width=1\linewidth]{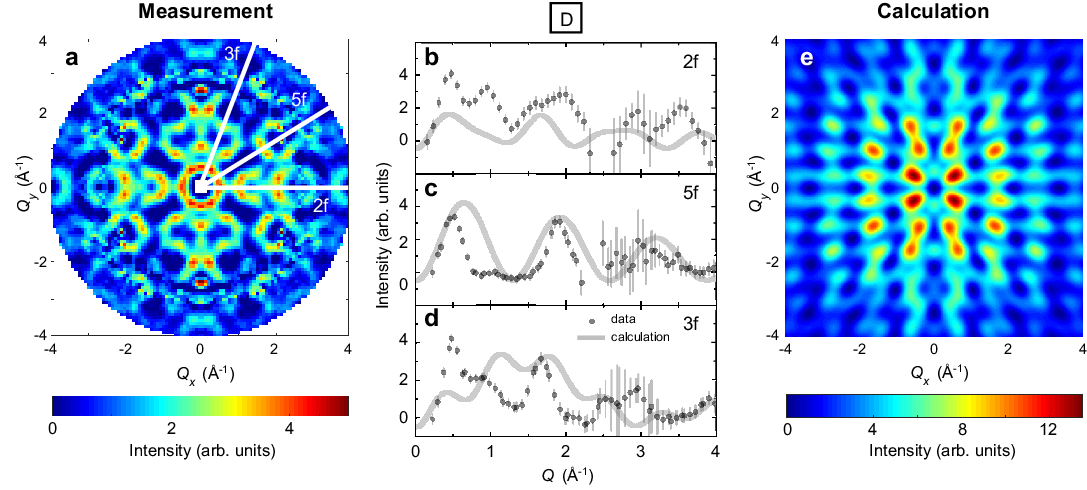}
  \caption{\label{FigA4}
           (Color online) Comparison of scans through the averaged magnetic diffuse scattering for Tb$^{3+}$ icosahedron moment configurations corresponding to region D of Figure \ref{Fig3}\\.}
\end{figure}
\clearpage

\begin{figure}
  \centering
  \includegraphics[width=1\linewidth]{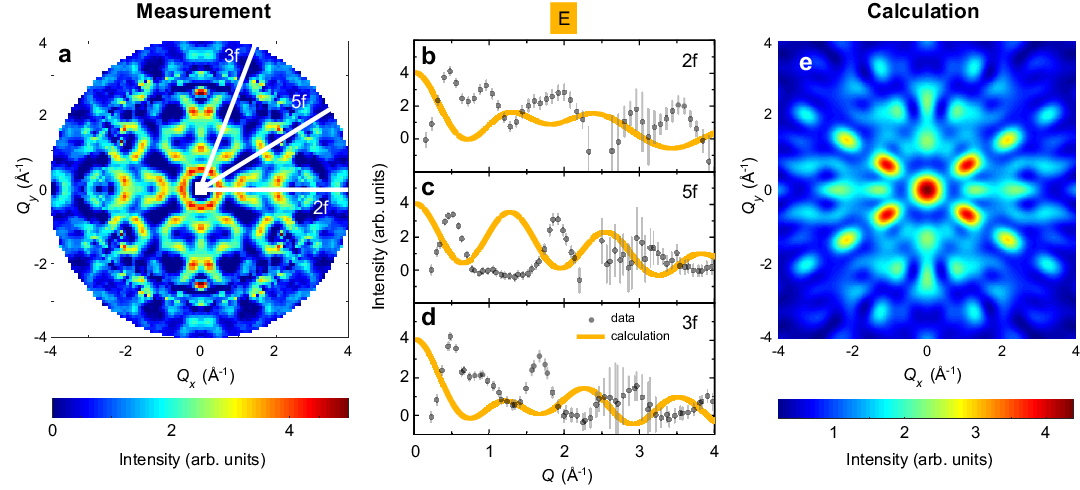}
  \caption{\label{FigA5}
           (Color online) Comparison of scans through the averaged magnetic diffuse scattering for Tb$^{3+}$ icosahedron moment configurations corresponding to region E of Figure \ref{Fig3}.\\}
\end{figure}

\begin{figure}
  \centering
  \includegraphics[width=1\linewidth]{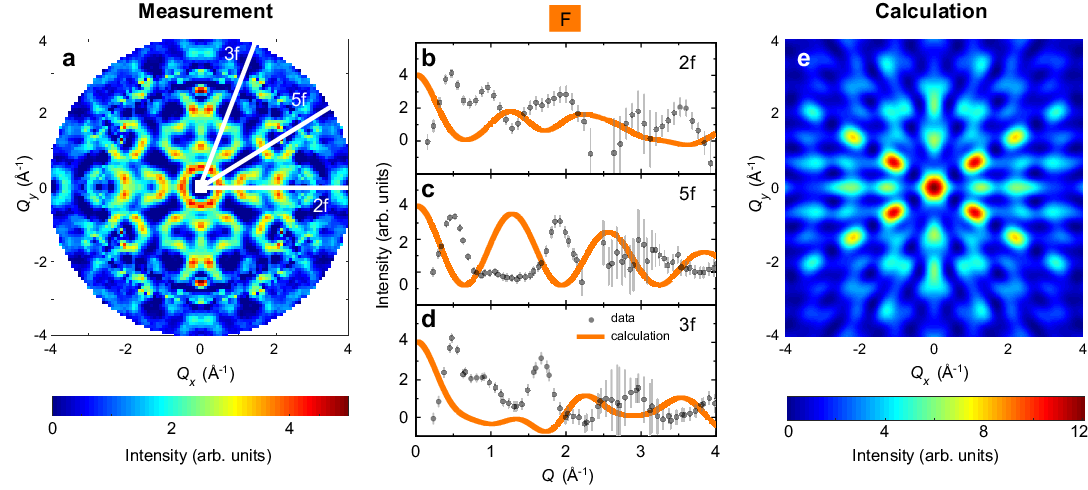}
  \caption{\label{FigA6}
           (Color online) Comparison of scans through the averaged magnetic diffuse scattering for Tb$^{3+}$ icosahedron moment configurations corresponding to region F of Figure \ref{Fig3}.\\}
\end{figure}
\end{widetext}

\end{document}